\documentclass[11pt]{article}
\usepackage{graphicx}
\renewcommand{\baselinestretch}{1}
\begin{document}
\title{\bf  Capability of the free-ion eigenstates for crystal-field splitting}
\author{\bf J. Mulak$^{1}$, M. Mulak$^{2}$}
\date{{\it  $^{1}$ Trzebiatowski Institute of Low Temperature
            and Structure Research,\\
            Polish Academy of Sciences, 50--950, PO Box 1410,
            Wroclaw, Poland\\
            $^{2}$ Institute of Physics,
            Wroclaw University of Technology,\\
            Wyb. Wyspianskiego 27,
            50--370 Wroclaw, Poland}}
\maketitle
\vspace*{0.2cm}
\noindent
{\bf Corresponding author:}\\
Prof. Jacek Mulak\\
Trzebiatowski Institute of Low Temperature and
Structure Research,\\
Polish Academy of Sciences,
50--950, PO Box 1410,
Wroclaw, POLAND\\
email: Maciej.Mulak@pwr.wroc.pl\\
Tel: (+4871) 3435021, 3443206\\
Fax: (+4871) 3441029\\
\begin{abstract}
\noindent Any electron eigenstate $|\Psi\rangle$ of the paramagnetic
ion open-shell is characterized by the three independent multipole
asphericities $A_{k}=\langle \Psi || C^{(k)}||\Psi\rangle$ for
$k=2,4$ and $6$ related to the second moments of the relevant
crystal-field splittings by
$\sigma_{k}^{2}=\left[1/(2J+1)\right]A_{k}^{2}S_{k}^{2}$ , where
$S_{k}^{2}=\left[1/(2k+1)\right]\sum_{q}|B_{kq}|^{2}$. The $A_{k}$
can serve as a reliable measure of the state $|\Psi\rangle$
capability for the splitting produced by the $k$-rank component of
the crystal-field Hamiltonian and allow one to verify any fitted
crystal-field parameter set directly comparing the calculated and
the experimental second moments of the relevant crystal-field
splittings. We present the multipole characteristics $A_{k}$ for the
extensive set of eigenstates from the lower parts of energy spectra
of the tripositive $4f^{N}$ ions applying in the calculations the
advanced eigenfunctions of the free lanthanide ions obtained based
on the M. Reid $f$-shell programs. Such amended asphericities are
compared with those achieved for the simplified Russell-Saunders
states. Next, they are classified with respect to the absolute or
relative weight of the $A_{k}$ in the multipole structure of the
considered states. For the majority of the analyzed states (about
80\%) the $A_{k}$ variation is of order of only a few percent. Some
essential changes are found primarily for several states of
Tm$^{3+}$, Er$^{3+}$, Nd$^{3+}$, and Pr$^{3+}$ ions. The detailed
mechanisms of such $A_{k}$ changes are unveiled. Particularly,
certain noteworthy cancelations as well as enhancements of their
magnitudes are explained.
\end{abstract}
\noindent
{\it PACS}: 71.15.-m, 71.23.An, 71.70.Ch, 75.10.Dg \\

\section*{1. Introduction}
The spherical harmonic operators $C_{q}^{(k)}$ in the one-electron
crystal-field (CF) Hamiltonian  ${\cal
H}_{CF}=\sum_{i}\sum_{k,q}B_{kq}C_{q}^{(k)}(\vartheta_{i},\varphi_{i})$
[1], written shortly as  ${\cal
H}_{CF}=\sum_{k,q}B_{kq}C_{q}^{(k)}$, act on the angle coordinates
$\vartheta_{i},\varphi_{i}$ of individual unpaired electrons $(i)$
of the central ion in its initial eigenstates $|\Psi\rangle$ that
are superpositions of the Russel-Saunders (RS) states
$|l^{N}SLJM_{J}\rangle$ . The $B_{kq}$ stand for the crystal-field
parameters (CFP) for the above specified operators. For complex
many-electron states the one-electron character of the $C_{q}^{(k)}$
operators manifests itself by the 6-$j$ symbols in their developed
matrix elements [1-5] and the doubly reduced matrix elements of the
unit tensor operator $U^{(k)}$ [1-5] (Section 2, Eq.(2)). They both
reveal a decomposition of the coupled many-electron state into its
one-electron spinorbitals. Thus, any matrix element $\langle \Psi
|C_{q}^{(k)}|\Psi\rangle$ is concerned exclusively with the
intrinsic properties of the central ion electron eigenstate
$|\Psi\rangle$. The reduced (double bar) matrix elements is defined
by [1-5]
\begin{equation}\label{1}
    \langle l^{N}SLJ||C^{(k)}||l^{N}SL^{\prime}J^{\prime}\rangle=
    (-1)^{J-M_{J}}\left(              \begin{array}{ccc}
                                         J & k & J^{\prime} \\
                                         -M_{J} & q & M_{J^{\prime}}
                                       \end{array}
    \right)^{-1}
    \langle
    l^{N}SLJM_{J}|C_{q}^{(k)}|l^{N}SL^{\prime}J^{\prime}M_{J^{\prime}}\rangle,
\end{equation}
where the factor preceding the matrix element is the reciprocal of
the 3-$j$ symbol [1-5]. The reduced matrix element is independent of
the reference frame orientation and hence also of $M_{J}$. The
diagonal reduced elements $\langle \Psi || C^{(k)}||\Psi\rangle$
represent the $2^{k}$-pole asphericities $A_{k}$ (for $k=2,4$ and
$6$) of the considered electron state $|\Psi\rangle$ [6] and these
dimensionless values can serve as a reliable measure of the state
capability for CF splitting by the $k$-rank CF Hamiltonian (Section
3). The electron density distribution of the $f$-electron states is
fully described by the first three multipoles with even $k=2,4$ and
$6$. The asphericities $A_{k}$ for 105 lower lying electron
eigenstates of all the trivalent lanthanide ions are compiled in
Table 1 (Section 2). They have been calculated for the corrected
eigenstates including $J$-mixing effect [7] and the outstanding set
of the free-ion data [8] and subsequently compared with those
corresponding to the one-component RS states [6]. The $A_{k}$
magnitudes and their possible variations due to the $J$-mixing of
the RS states are thoroughly discussed. An inseparable entanglement
of the asphericities $A_{k}$ and the $k$-components of the CF
strength $S_{k}$ [9-12] seen in the expression for the second moment
of the splitting $\sigma_{k}$ [10, 13, 14] (Section 3) justifies the
$A_{k}$ as a reliable capability of the relevant state for the
$2^{(k)}$-pole partial CF splitting. By the fundamental law of
additivity $\sigma^{2}=\sum_{k}\sigma^{2}_{k}$ [10, 13, 14],
resulting from the orthogonality of the $3$-j symbols [3,4]
(Eq.(1)), the global $\sigma^{2}$ can be expressed by means of the
$A_{k}$ and $S_{k}$ components. Tables 2-6 show the classification
of the examined eigenstates with respect to their multipole
structure (Section 4). The states distinguished by the strongest and
the weakest $A=(A_{2}^{2}+A_{4}^{2}+A_{6}^{2})^{1/2}$, by the
strongest and the weakest $|A_{k}|$, and finally those with the
largest and the smallest $|A_{k}|/A$ for $k=2,4$ and $6$ have been
selected respectively. In turn, Section 5 gives a few instructive
examples unveiling the mechanisms of the $A_{k}$ changes induced by
the $J$-mixing of the RS states. A special attention has been paid
to the strong enhancements and cancelations among the asphericities
$A_{k}$.

\section*{2. Multipole characteristics of the $4f^{N}$ tripositive free-ion eigenstates including $J$-mixing effects}
The $k$-rank multipole moment of an electron eigenstate
$|\Psi\rangle$ which is a superposition of the RS states with
various $L$ and $S$ but the same $J$ can be evaluated based on the
reduced matrix element $\langle \Psi || C^{(k)}||\Psi\rangle$ of the
respective $k$-rank spherical harmonic operator. According to the
Wigner-Eckart theorem [5,15] such quantity is independent of the
reference frame orientation and adequately expresses the
$2^{k}$-pole type asphericity of the given eigenstate
$|\Psi\rangle$. For the spherical electron density distribution the
matrix element identically vanishes for $k=2,4$ and $6$. It plays
also a crucial role as a scaling factor in the CF Hamiltonian
interaction matrices and hence participates in both the
calculational and fitting CFP procedures. In the case of $J$-mixing
approach, i.e. for fixed $J$, the reduced matrix element can be
expressed by the sum of all diagonal and off-diagonal matrix
elements occurring in the $\langle \Psi || C^{(k)}||\Psi\rangle$
expansion [1,5,16]
\begin{equation}\label{2}
    \langle l^{N}SLJ||C^{(k)}||l^{N}SL^{\prime}J\rangle=
    (-1)^{S+L^{\prime}+J+k}(2J+1)\left\{              \begin{array}{ccc}
                                         J & J & k \\
                                         L^{\prime} & L & S
                                       \end{array}
    \right\}
    \langle
    l^{N}SL||U^{(k)}||l^{N}SL^{\prime}\rangle \langle l || C^{(k)}||l \rangle,
\end{equation}
where the first factor on the right side defining the sign of the
reduced element depends on the parity of the sum of four numbers, in
principle autonomous, what leads to the sign randomness. The second
factor stands for the degeneracy of the state, the third one is the
$6$-j symbol revealing what part of the final $|SLJ\rangle$ function
belongs to the orbital part $|SL^{\prime}\rangle$ [17]. Finally, the
double-bar matrix element of the unit tensor operator $U^{(k)}$
depends how $N$ one-electron angular momenta
$\vec{\textbf{\emph{l}}}$ of the $l^{N}$ configuration couple into
the resultant $\vec{\textbf{\emph{L}}}$ [18]. The one-electron
reduced matrix element $\langle l || C^{(k)}||l \rangle$ for $l=3$
is equal to $-1.3663$, $1.1282$ and $-1.2774$ for $k=2,4$ and $6$,
respectively.

The $M_{J}$ quantum numbers and the $q$ index do not appear in
Eq.(2) (compare with Eq.(1)). It clearly shows that the reduced
matrix elements and in consequence the $A_{k}$ are independent of
the reference frame choice. Any element of the $\langle \Psi ||
C^{(k)}||\Psi \rangle$ expansion encloses additionally the product
of amplitudes of the two involved components in the $|\Psi\rangle$
superposition together with their signs. The reduced matrix element
(Eq.(2)) differs from zero only for the same $S$ quantum number (in
the bra and ket) since the $C^{(k)}$ acts exclusively on the
configurational coordinates of the electrons, and for the same
parity of $L$ and $L^{\prime}$. These requirements reduce the number
of the non-zero off-diagonal matrix elements between various
components of the $J$-mixed eigenfunctions.

Such multipole characteristics have been evaluated earlier for the
pure (one-component) RS open-shell electron eigenstates [6]. In
Table 1 we compare them with the corrected characteristics for the
$4f^{N}$ tripositive ion eigenstates obtained in the more accurate
$J$-mixing approach based on the M. Reid $f$-shell programs [7] and
the free-ion data reported by Carnall at al [8]. In the considered
$J$-mixed superpositions the average number of RS components amounts
to $7$, whereas the average number of the constituent matrix
elements $13$. In turn, the maximal number of the components reaches
$22$, whereas the maximal number of the matrix elements amounts to
$64$ (including $42$ off-diagonal ones) what occurs for the ninth
eigenstate of Dy$^{3+}$ ion (No.70 in Table 1) with
$|^{6}F_{7/2}\rangle$ state as the upper component.

In total, we have taken into account $105$ lower lying eigenstates
of the three-valent RE ions from Ce$^{3+}$ ($4f^{1}$) up to
Yb$^{3+}$ ($4f^{13}$). Table 1 covers also the basic attributes of
the considered eigenstates: the upper RS component, its amplitude in
the normalized superposition, the consecutive number in the ion's
spectrum [7], the eigenenergy in $cm^{-1}$, and the number of
components with the amplitude exceeding $0.01$. The number in the
first column of Table 1 identifies any quoted state or its
parameters consistently throughout the whole paper. It is
instructive to compare the asphericities of the pure RS states [6]
with those of the corrected $J$-mixed eigenstates. It turns out that
from among the 105 analysed states only about 20$\%$ of them differ
markedly in the asphericities from their RS counterparts, i.e. their
upper states. Primarily, these are the states of the following ions:
Tm$^{3+}$ ($4f^{12}$), Er$^{3+}$ ($4f^{11}$), Nd$^{3+}$ ($4f^{3}$),
and Pr$^{3+}$ ($4f^{2}$) (Table 1). By sheer coincidence two various
states of Tm$^{3+}$ ion: No.99 and No.103 are characterized by the
same dominating component $|^{3}P_{2}\rangle$, but it does not lead
to any misunderstanding because we do not use this ambiguous state
description.

There exist the following $J$-mixing mechanisms that produce the
observed changes in the asphericity of the states. Firstly, the
normalization of any superposition of states reduces naturally the
upper state amplitude, whereas its square determines the upper state
asphericity input. Secondly, additional diagonal and off-diagonal
terms in the the matrix element $\langle \Psi || C^{(k)}||\Psi
\rangle$ expansion differ in magnitudes and signs. The sign of each
individual diagonal term is specified exclusively by the sign of the
respective $A_{k}$ on the involved component. Its magnitude,
however, comes from the product of the $|A_{k}|$ and the square of
the component amplitude in the superposition. In turn, any
off-diagonal term is a product of 6 factors including two involved
amplitudes (Eq.(2)). Its sign results from the product of 6 signs,
in principle autonomous. To cope with this matter effectively one
should consider all the additional diagonal and off-diagonal
contributions along with their various possible magnitudes and
signs. Based on these investigations four types of the resultant
$A_{k}$ modifications can be noticed in Table 1:
\begin{itemize}
  \item [(i)] Due to insignificant $J$-mixing admixtures
  to the upper state only small changes (within a few
  percent) arise in the pertinet $|A_{k}|$, which are the algebraic
  sum of the normalization effect and the additional diagonal and
  off-diagonal corrections. Such effect occurs for about 80$\%$ of
  the states presented in Table 1. However, the proximity of the
  $A_{k}$ for the RS and the corrected $J$-mixed states can be
  also accidental. By way of example, for eigenstate No.28 of Nd$^{3+}$ ion
  the upper state $|^{3}P_{3/2}\rangle$ amplitude reaches merely
  0.7205 and its contribution to the $A_{2}$ of the superposition
  only $(0.7205)^{2}(0.2981)=0.1548$. Nevertheless, the remaining
  diagonal (0.2128) and off-diagonal (-0.0858) inputs,
  relatively large, lead in sum to the $A_{2}=0.2818$ that accidentally
  is close to 0.2981, which is the value for the $|^{3}P_{3/2}\rangle$
  state.
  \item [(ii)] The sum of the corrections is substantial with
  respect to the upper state $A_{k}$ and has the same sign as the
  $A_{k}$. Here an enhancement of the $|A_{k}|$ occurs. Such resultant
  effect is observed for the states No.: 27, 88, 97, 98, and 103 in
  the case of $A_{2}$, for the states No.: 8, 64, 65, and 93 in the case
  of $A_{4}$, and for the state No.95 in the case of $A_{6}$.
  \item [(iii)] The sum of the corrections is substantial but with
  the opposite sign than that of the upper state $A_{k}$. In this case a
  partial compensation of the $|A_{k}|$ (including the complete one)
  or even the sign conversion of $A_{k}$ takes place. Such result
  has been found in the case of $A_{2}$ for the states No.: 8, 9, 21,
  24, 25, 79, 84, 85, 86, 93, 95, 99, in the case of $A_{4}$ for the
  states No.: 9, 10, 21, 24, 25, 85, 86, 88, 95, 97, 98, and in
  the case of $A_{6}$ for the states No.: 3, 8, 9, 21, 24, 25, 27, 64,
  65, 79, 84, 85, 86, 88, 93, 98.
  \item [(iv)] The corrections generate the only contribution to
  $A_{k}$ that for the initial state is equal to zero. It takes
  place  for the states No.14 ($A_{4}$),  No.23 ($A_{2}$),  No.80 ($A_{2},
  A_{4}$),  No.87 ($A_{2}$),  No.99 ($A_{4}$),  No.103 ($A_{4}$).
\end{itemize}
The detailed mechanisms of the asphericity modifications induced by
the $J$-mixing effect will be thoroughly analyzed for some
representative examples in Section 5.

\section*{3. The asphericity of an electron eigenstate and its crystal-field splitting}
The asphericity $A_{k}$ for $k=2,4$ and $6$ of any electron state
may serve as a reliable measure of its capability for CF splitting
produced by the ${\cal H}_{CF}^{(k)}$ - the $k$-th component of the
${\cal H}_{CF}$. It stems from the fundamental relationship between
the CF splitting second moment $\sigma_{k}$ and the $A_{k}$
[10,13,14]
\begin{equation}\label{3}
    \sigma_{k}^{2}=\left[1/(2J+1)\right]A_{k}^{2}S_{k}^{2},
\end{equation}
where $S_{k}^{2}=\left[1/(2k+1)\right]\sum_{q}|B_{kq}|^{2}$ is the
square of the CF strength of the $2^{k}$-pole ${\cal H}_{CF}$
component [9-12], and $(2J+1)$ is the degeneracy of the given state
with a good quantum number $J$. In fact, the above relationship
(Eq.(3)) implies from the spherical harmonic addition theorem [19]
concerning the expansion of $1/r_{ij}$ into the series of
$C_{q}^{(k)}(\vartheta_{i},\varphi_{i})\cdot
C_{q}^{(k)\ast}(\theta_{j},\phi_{j})$ components. They are the
products of the conjugated spherical harmonics defined for the
separated indices $i$ and $j$. In the CF context the first factor
refers to the electron density angle distribution of the central ion
unperturbed eigenstate, whereas the second refers to the surrounding
charges. In fact, this separation lies in the background of the
whole formalism.

As is seen from Eq.(3) the asphericity $A_{k}$ can be treated as a
potential capability of the considered state for the $2^{k}$-pole CF
splitting since the second factor $S_{k}$ represents a separate and
unrelated external impact. The $A_{k}$ can be either positive or
negative (Section 2) what symbolically may be imagined as
asphericities of convex or concave type. The $A_{k}$ sign is
irrelevant for the $\sigma_{k}$, but is crucial calculating the
resultant asphericities of the superposition of states.

The question arises how the global second moment $\sigma$ can be
expressed by means of the asphericities of the involved electron
eigenstate. As is known, the square of the global second moment
$\sigma^{2}$ is a simple sum of the second moment squares of the
individual components.

\begin{equation}\label{4}
    \sigma^{2}=\left[1/(2J+1)\right](A_{2}^{2}S_{2}^{2}+A_{4}^{2}S_{4}^{2}+A_{6}^{2}S_{6}^{2}).
\end{equation}
To describe $\sigma^{2}$ it is convenient to introduce two auxiliary
and figurative vectors: ${\cal A} (A_{2}^{2},A_{4}^{2},A_{6}^{2})$
and ${\cal S} (S_{2}^{2},S_{4}^{2},S_{6}^{2})$ within the
three-dimensional orthogonal reference frame based on the
$A_{k}^{2}$ (or $S_{k}^{2}$) axes. Then,
$\sigma^{2}=\left[1/(2J+1)\right] {\cal A}\cdot{\cal S}$ is defined
by their scalar product. All the components of the ${\cal A}$ and
${\cal S}$ vectors are positive by definition and can be expressed
by the spherical angle coordinates only within the ranges of
$0\leq\theta\leq \pi/2$ and $0\leq\phi\leq \pi/2$. Eq.(4) shows that
the CF splitting is determined by the two inseparable independent
quantities $A_{k}$ and $S_{k}$ mutually entangled. The figurative
vectors ${\cal A}$ and ${\cal S}$ may be orthogonal, what happens
when both the vectors lie either on the two frame axes or one of
them lies along an axis whereas the second belongs to the
perpendicular plane. Then, always $\sigma^{2}=0$ in spite of some
non-zero $A_{k}$ and $S_{k}$. Simultaneously, Eq.(4) enables us to
critically verify the meaning of such quantities like $S=\left(
\sum_{k}S_{k}^{2}\right)^{1/2}$ and $A=\left(
\sum_{k}A_{k}^{2}\right)^{1/2}$ [6]. In general, no apparent
physical sense can be assigned to them.

\section*{4. The range of capability of the $4f^{N}$ tripositive free ion eigenstates for crystal-field splitting}
Similarly to the approximated RS states $|^{2S+1}L_{J}\rangle$ of
triply ionized lanthanides [6], the eigenstates amended by the
$J$-mixing [7] are characterized by an exceedingly diversified
multipole structure both in qualitative and quantitative way (Table
1). Such random to a large extent diversity stems from a stochastic
character with respect to the magnitude and sign of the
multifactorial expression for the $C^{(k)}$ operator reduced matrix
element (Eq.(2)). The chaotic dispersion of the $A_{k}$ magnitudes
and signs is well exhibited in Tables 2-6 by the eigenstates chosen
from among all the 105 studied ones: the top ten states of the
strongest or weakest $A=(A_{2}^{2}+A_{4}^{2}+A_{6}^{2})^{1/2}$
(Table 2), the ten ones of the strongest $|A_{k}|$ (Tables 3), the
ten of the weakest $|A_{k}|$ (Table 4), and finally the ten states
of the highest $|A_{k}|/A$ (Table 5), as well as the ten ones of the
lowest $|A_{k}|/A$ (Table 6). The $|A_{k}|/A$, which is a cosine of
the angle between the $(A_{2},A_{4},A_{6})$ vector and the
distinguished axis representing the $A_{k}$, gives the relative
weight of the chosen $2^{k}$-pole in the eigenstate multipole
structure. It is enough to notice that $A$ takes values from 0 to
3.3784, whereas the entirely independent one of another $|A_{k}|$
change within the ranges: $0<|A_{2}|\leq 3.0273$, $\;0<|A_{4}|\leq
1.4303$, and $0<|A_{6}|\leq 1.7765$. In consequence, the multipole
structure of the considered states is widely differentiated and can
assume also the forms with only one prevailing multipole. For
example, No.14 eigenstate $|^{3}P_{2}\rangle$ of Pr$^{3+}$ ion is
characterized by the predominating role of the $2^{2}$-pole
component $(|A_{2}|/A)=0.9917)$, No.43 eigenstate
$|^{6}F_{5/2}\rangle$ of Sm$^{3+}$ ion by the $2^{4}$-pole component
$(|A_{4}|/A)\simeq 1.0000)$, and No.18 eigenstate
$|^{4}I_{15/2}\rangle$ of Nd$^{3+}$ ion by the prevailing
$2^{6}$-pole component $(|A_{6}|/A)=0.8812)$, however not so
distinctly as in the two previous cases.

The highest total asphericities (the top $A$ values), it means the
strongest total capabilities for the CF splitting, are found in the
states with large $L$ (and $J$) quantum numbers (Table 2). Such
states are weakly disturbed by the $J$-mixing interaction due to a
small number of the partner RS states of the same $J$ and large
energy gaps between them. Their calculated asphericities are close
to those for the relevant upper states. On the contrary, the
eigenstates with the weakest asphericities have quite often  their
$A_{k}$ significantly changed with respect to those for their RS
counterparts. In general, it results from a similar level of the
$J$-mixing corrections in both the cases, and a substantial
difference in their initial magnitudes.

Tables 1-6 enable to note an evident correspondence between the
calculated $A_{k}$ for the pairs of the lanthanide ions with the
complementary electron configurations $4f^{N}$ and $4f^{14-N}$:
(Ce$^{3+}$, Yb$^{3+}$), (Pr$^{3+}$, Tm$^{3+}$), (Nd$^{3+}$,
Er$^{3+}$), (Pm$^{3+}$, Ho$^{3+}$), (Sm$^{3+}$, Dy$^{3+}$) and
(Eu$^{3+}$, Tb$^{3+}$). The opposite $A_{k}$ sign of the
pair-partners results from the opposite sign of the related matrix
elements of the $U^{(k)}$ operators [18], and is mainly a
repercussion of the Hund's rules governing the eigenstates sequence,
it means their location in the free-ion energy spectrum. As a good
example of such case may serve the difference between the bottom
parts of the energy diagrams of Pr$^{3+}$ and Tm$^{3+}$ ions. In the
energy spectrum of Pr$^{3+}$ ion the RS states $|^{3}H_{4}\rangle$
and $|^{3}F_{4}\rangle$ interacting via $J$-mixing are located one
to anther as far as possible: the $|^{3}H_{4}\rangle$ is the lowest
state of the term $^{3}H$, whereas the $|^{3}F_{4}\rangle$ the
highest one of the term $^{3}F$. In Tm$^{3+}$ ion, in the reverse
order, the $|^{3}H_{4}\rangle$ is the highest state of the $^{3}H$
term, whereas the $|^{3}F_{4}\rangle$ the lowest state of the term
$^{3}F$. In fact, the $|^{3}F_{4}\rangle$ state lies below the state
$|^{3}H_{4}\rangle$ [16]. The energy gap between the states
$|^{3}H_{4}\rangle$ and $|^{3}F_{4}\rangle$, their so-called energy
denominator, determines the efficiency of the $J$-mixing
interaction.

\section*{5. Discussion }
The calculated asphericities $A_{k}=\langle \Psi ||
C^{(k)}||\Psi\rangle$ of the trivalent $4f^{N}$ ions are not the
actual ones due to approximate nature of the applied eigenfunctions
$|\Psi\rangle$, but their reliability can be improved replacing the
initial functions (e.g. those of the RS type) by their various
superpositions. In the case of simultaneous diagonalisation of the
interaction matrix including the Coulomb repulsion and the
spin-orbit coupling these are the superpositions of the RS functions
with the same $J$ but different $L$ and $S$ quantum numbers [7]. The
$A_{k}$ variations seen in Table 1 are limited mainly by the scale
of the component admixtures. Additional role play the magnitudes of
the relevant diagonal and off-diagonal matrix elements of the
$C^{(k)}$ operator within the superposition, as well as the mutual
competition between the corrections. In most cases the amplitudes of
the admixtures are rather small. Therefore, for the majority of the
lower lying eigenstates (about 80\%) of the trivalent lanthanide
ions there appear only inconsiderable differences between the
$A_{k}$ calculated for the model RS states [6] and those including
their $J$-mixing (Table 1). Nevertheless, for certain part of the
eigenstates, particularly the exited ones, the observed changes
become essential, indeed. They illustrate well the types of the
resultant $J$-mixing effects mentioned in Section 2. Some
instructive mechanisms leading to such variations are analyzed in
details below for several chosen examples.

Let us consider No.8 eigenstate in Table 1. It is the sixth state of
Pr$^{3+}$ ion of the composition:
$0.8087|^{3}F_{4}\rangle+0.1225|^{3}H_{4}\rangle-0.5753|^{1}G_{4}\rangle$
with the dominating $|^{3}F_{4}\rangle$ component. The diagonal
contributions to the $A_{2}$ amount to:
$(0.8087)^{2}(0.4672)=0.3778$, $(0.1225)^{2}(-1.2367)=-0.0186$,
$(-0.5753)^{2}(-0.3058)=-0.1012$, and the only off-diagonal input
$\langle(0.8087)\;^{3}F_{4}||C^{(2)}||(0.1225)\;^{3}H_{4}\rangle=-0.0141$.
The accumulation of the three negative corrections reduces the
$A_{2}$ from $0.4672$ down to $0.2439$. The diagonal contributions
to the $A_{4}$ are negative and reach:
$(0.8087)^{2}(-0.2906)=-0.1901$, $(0.1225)^{2}(-0.7395)=-0.0111$,
$(-0.5753)^{2}(-1.2150)=-0.4021$, and the off-diagonal element
$\langle(0.8087)\;^{3}F_{4}||C^{(4)}||(0.1225)\;^{3}H_{4}\rangle=0.0549$.
Here, the strong diagonal input of the $|^{1}G_{4}\rangle$ governs
the magnitude and sign of the $A_{4}=-0.2906$. In turn, the diagonal
contributions to the $A_{6}$ are equal to:
$(0.8087)^{2}(0.1558)=0.1019$, $(0.1225)^{2}(0.7706)=0.0116$,
$(-0.5753)^{2}(-1.5299)=-0.5064$, and the off-diagonal one
$\langle(0.8087)\;^{3}F_{4}||C^{(6)}||(0.1225)\;^{3}H_{4}\rangle=0.1649$.
Again, as above, the diagonal negative input of the
$|^{1}G_{4}\rangle$ dominates and the ultimate $A_{6}=-0.2280$
results from a partial compensation of all the contributions.

Eigenstate No.21, the seventh state of Nd$^{3+}$ ion, is composed of
$-0.3700|^{4}F_{9/2}\rangle-0.1458|^{4}G_{9/2}\rangle+
0.1525|^{4}I_{9/2}\rangle+0.3381|^{2}G(1)_{9/2}\rangle-
0.2799|^{2}G(2)_{9/2}\rangle
-0.2805|^{2}H(1)_{9/2}\rangle+0.7398|^{2}H(2)_{9/2}\rangle\;$ with
the prevailing $|^{2}H(2)_{9/2}\rangle$ state. All the weak diagonal
contributions to the $A_{2}$ are almost compensated achieving in sum
0.0092 with respect to the dominating state input
$(0.7398)^{2}(-0.0069)=-0.0038$. The decisive are the positive
off-diagonal terms
$\langle(0.3381)\;^{2}G(1)_{9/2}||C^{(2)}||(-0.2799)^{2}G(2)_{9/2}\rangle=0.1655$,
and
$\langle(0.2805)\;^{2}H(1)_{9/2}||C^{(2)}||(0.7398)\;^{2}H(2)_{9/2}\rangle=0.1225$,
giving finally the $A_{2}=0.2920$. Here, the dominating state input
to the $A_{4}$ amounts to $(0.7398)^{2}(0.4816)=0.2636$ and the sum
of all the seven diagonal elements $0.1675$ is somewhat less. In
this situation the relatively large and negative off-diagonal
element
$\langle(0.2805)\;^{2}H(1)_{9/2}||C^{(4)}||(0.7398)\;^{2}H(2)_{9/2}\rangle=-0.2508$
decides both on the magnitude and sign of the $A_{4}=-0.0638$.
Similarly, for the very small positive sum of the partial diagonal
elements $(0.0055)$, the final $A_{6}=-0.0732$ is determined by the
prevailing, as for the modulus, negative off-diagonal element
$\langle(0.2805)\;^{2}H(1)_{9/2}||C^{(6)}||(0.7398)\;^{2}H(2)_{9/2}\rangle=-0.1010$.

Eigenstate No.25, the eleventh state of the Nd$^{3+}$, is given by
$-0.2407|^{4}G_{11/2}\rangle+0.0994|^{4}I_{11/2}\rangle-0.3573|^{2}H(1)_{11/2}\rangle+
0.8955|^{2}H(2)_{11/2}\rangle-0.0515|^{2}I_{11/2}\rangle$  with the
dominating $|^{2}H(2)_{11/2}\rangle$ component. The sum of the
diagonal contributions to the $A_{2}$ is -0.0740, including the
input -0.0632 from the $|^{2}H(1)_{11/2}\rangle$. The resultant
$A_{2}=0.1289$ is the outcome of mutual competition of the positive
off-diagonal term
$\langle(-0.3573)\;^{2}H(1)_{11/2}||C^{(2)}||(0.8955)\;^{2}H(2)_{11/2}\rangle=0.2080$
and the negative diagonal contribution coming mainly from the state
$|^{2}H(1)_{11/2}\rangle$. The sum of the diagonal elements
combining to the $A_{4}$ amounts to 0.4454 and is close to the
contribution of the dominating $|^{2}H(2)_{11/2}\rangle$ state, i.e.
$(0.8955)^{2}(0.5373)=0.4309$. However, it is practically entirely
compensated $(A_{4}=0.0001)$ by the sum of two negative off-diagonal
elements:
$\langle(-0.3573)\;^{2}H(1)_{11/2}||C^{(4)}||(0.8955)\;^{2}H(2)_{11/2}\rangle=-0.4314$
and
$\langle(-0.2407)\;^{4}G_{11/2}||C^{(4)}||(0.0994)\;^{4}I_{11/2}\rangle=-0.0139$.
The resultant $A_{6}=-0.2027$ is determined by relatively strong
off-diagonal input
$\langle(-0.3573)\;^{2}H(1)_{11/2}||C^{(6)}||(0.8955)\;^{2}H(2)_{11/2}\rangle=-0.1791$.
All the diagonal elements contribute only -0.0081.

The $J$-mixing of the RS states can activate some idle states making
them susceptible to CF splittings. In other words, they lose their
initial effective spherical symmetry. As an example let us examine
eigenstate No.87, the sixth state od Er$^{3+}$ ion consisting of
$0.8293|^{4}S_{3/2}\rangle+0.0443|^{4}D_{3/2}\rangle+0.2390|^{4}F_{3/2}\rangle-
0.4174|^{2}P_{3/2}\rangle-0.2797|^{2}D(1)_{3/2}\rangle-0.0274|^{2}D(2)_{3/2}\rangle$.
The prevailing element $|^{4}S_{3/2}\rangle$ is characterized by
zero asphericities $A_{2}$, $A_{4}$ and $A_{6}$. However, the
corrected eigenstate acquires the asphericity $A_{2}=-0.1689$ by
accumulation of the negative diagonal contributions:
$(0.2390)^{2}(-0.3578)=-0.0204$,  $(-0.4174)^{2}(-0.2981)=-0.0519$,
$(-0.2797)^{2}(-0.5707)=-0.0446$, and the off-diagonal ones:
$\langle(0.0443)\;^{4}D_{3/2}||C^{(2)}||(0.8293)\;^{4}S_{3/2}\rangle=-0.0480$,
$\langle(-0.2797)\;^{2}D(1)_{3/2}||C^{(2)}||(-0.0274)\;^{2}D(2)_{3/2}\rangle=-0.0045$.
The states $|^{4}S_{3/2}\rangle$ and $|^{4}D_{3/2}\rangle$ do not
bring any diagonal inputs, and the state $|^{2}D(2)_{3/2}\rangle$
gives only 0.0005.

In the case of eigenstate No.3 the ground state of Pr$^{3+}$ ion is
given by
$0.9856|^{3}H_{4}\rangle+0.1662|^{1}G_{4}\rangle-0.0311|^{3}F_{4}\rangle$
and its $A_{2}$ and $A_{4}$ asphericities change only slightly with
respect to the parameters for the pure  $|^{3}H_{4}\rangle$ state.
However, the $|A_{6}|$ asphericity is noticeably reduced. The
diagonal contribution of the $|^{1}G_{4}\rangle$ state
$(0.1662)^{2}(-1.5299)=-0.0423$, and the off-diagonal term which is
equal to
$\langle(0.9856)\;^{3}H_{4}||C^{(6)}||(-0.0311)\;^{3}F_{4}\rangle=-0.0510$
weaken the positive input of the $|^{3}H_{4}\rangle$ upper state
$(0.9856)^{2}(0.7706)=0.7486$ down to the value of 0.6555. It
corresponds to attenuation of the state capability for the CF
splitting by ${\cal H}_{CF}^{(6)}$. An increase in both the
$|^{3}F_{4}\rangle$ and $|^{1}G_{4}\rangle$ admixtures deepen the
tendency. It is worth to remember analyzing the CF splitting of the
U$^{4+}(5f^{2})$ ion ground state.

As is seen in Eq.(3) the multipole characteristics of the electron
eigenstates along with their CF splitting diagrams sheds a new light
on the crystal matrix multipole structure and vice versa. Based on
the CF splitting diagrams for several electron eigenstates of known
multipole characteristics in a specified crystal matrix (with a
definite ${\cal H}_{CF}$), as well as the CF splitting diagrams of a
specified eigenstate in various CF matrices, we can reconcile the
actual $A_{k}$ for the considered electron states and the $S_{k}$
for the ${\cal H}_{CF}$s, respectively. A great facilitation in such
estimations is an incomplete multipolar structure of the analyzed
eigenstates. Such incompleteness may result either from the triangle
rule for $J$, $J$, $k$ numbers (e.g. for $J=1,3/2$ and $k=4,6$ or
$J=5/2$ and $k=6$) or from accidental cancelation of some multipoles
due to the $J$-mixing effect, as it is observed for eigenstates
No.25 and No.85 in Table 1. Furthermore, in some CF Hamiltonians the
three-component multipole structure is not always complete, like
e.g. in the cubic ${\cal H}_{CF}$ which is deprived of quadrupolar
moment.

The reliable classification of the multipolar structure of both the
electron eigenstates and the actual CF Hamiltonians need such kind
of comprehensive investigations. The electron eigenstates
asphericities $A_{k}$ allows us to verify the fitted CFP sets
comparing the calculated (Eq.(3)) and experimental second moments of
the splitting [20]. The fitted CFP sets that well reproduce the
experimental spectrum of energy levels for intentionally
approximated initial eigenfunctions have by definition an effective
character. Therefore, applying the same approximation for all
eigenfunctions coming from different energy ranges will undoubtedly
lead to errors. Presumably, this is the main reason for difficulties
associated with minimization of $rms$ deviations in fitted CFP sets.
There are some phenomenological attempts to improve the fitting
accuracy. In one of them the two-electron correlation CF is
introduced which may be simple expressed by an effective
one-electron CF Hamiltonian being dependent on the electron term. In
another one the mean $k$ powers of the unpaired electron radii
$\langle r^{k}\rangle$, is made variable with respect to the
electron term [5,21,22]. Both the above approaches are formally
admissible, but they can be physically ungrounded.

Yet another reflection arises. The dichotomic structure of the CF
Hamiltonian [6] and random diversity of the asphericities do not
entitle us by no means to exploit the concept of the ${\cal H}_{CF}$
multipole series convergence, as the multipole series of an external
potential only. The ${\cal H}_{CF}$ approximation reducing its
multipole structure only to first quadrupolar term is groundless. An
exception could be perhaps a unique case when $A_{2}=A_{4}=A_{6}$.
Obviously, the ${\cal H}_{CF}$ three-multipole $(k=2,4,6)$ series is
a finite one, and not truncated. The higher multipoles do not
contribute at all. The second independent factor that controls to a
similar extent as the external multipoles the resultant hierarchy of
the three CF Hamiltonian terms is the capability $A_{k}$ of the
state for the CF splitting.
\clearpage


\renewcommand{\baselinestretch}{1}
\begin{table*}[b]
  \caption{Multipole character of the J-mixed electron eigenstates $|\Psi\rangle$ in RE free ions.
  The eigenfunctions and eigenvalues are calculated using M. Reid f-shell programs [7] and
  free-ion data reported by Carnall et al [8].
  The electron eigenstate data cover respectively: the upper component
  (Up.Comp.), its amplitude (Ampl.), consecutive no. in the spectrum
  [7] (no.), energy $[cm^{-1}]$ (E), number of components of amplitude $>0.01$
  (n). The multipolar asphericities for the upper
  component of the state are given in the parentheses.}
  \begin{center}
\begin{tabular}[htbp]{@{}cccclcrcccc@{}}
&&&&&&&&& \\
 \hline \textbf{No}.    &  \multicolumn{2}{c}{\textbf{RE
ion}} &\multicolumn{5}{c}{\textbf{Electron eigenstate data}}
    &  \multicolumn{3}{c}{\textbf{Multipolar asphericity}} \\
    & Ion          &     conf. & Up.Comp.
    & Ampl. & no.& E & n  & $A_{2}(\Psi)$   &  $A_{4}(\Psi)$  & $A_{6}(\Psi)$  \\
\hline
\textbf{1} & Ce$^{3+}$ & $f^{1}$ &$|^{2}F_{5/2}\rangle$ & 1 & 1 & 0 &    1          &  -1.1711  &   0.7560   &  0  \\
           &           &         &                      &   &   &
           &           &  (-1.1711)      &       (0.7560)                & (0)\\
\hline
\textbf{2} & Ce$^{3+}$ & $f^{1}$ &$|^{2}F_{7/2}\rangle$ & 1 & 2 & 2750 & 1          &  -1.3801 &   0.9670   &  -0.3054  \\
           &           &         &                      &   &   &
           &           &  (-1.3801)      &       (0.9670)               & (-0.3054 ) \\
\hline
\textbf{3} & Pr$^{3+}$& $f^{2}$ &$|^{3}H_{4}\rangle$ & 0.9856 & 1 & 0 & 3           &  -1.2048  &   -0.7693  &  0.6555    \\
\hline
  \end{tabular}
\end{center}
\end{table*}
\noindent

\clearpage

\begin{table*}[b]
\noindent Table 1 - cont. \\
\begin{center}
  \begin{tabular}[htbp]{@{}cccclcrcccc@{}}
\hline \textbf{No}.    &  \multicolumn{2}{c}{\textbf{RE ion}}
&\multicolumn{5}{c}{\textbf{Electron eigenstate data}}
    &  \multicolumn{3}{c}{\textbf{Multipolar asphericity}} \\
    & Ion          &     conf. & Up.Comp.
    & Ampl. & no.& E & n  & $A_{2}(\Psi)$   &  $A_{4}(\Psi)$  & $A_{6}(\Psi)$  \\
\hline
\textbf{21} & Nd$^{3+}$ & $f^{3}$       &$|^{2}H(2)_{9/2}\rangle$    & 0.7398    & 7     & 12519   & 7        &  0.2920  &  -0.0638  &  -0.0732  \\
  &           &         &                       & & & &                                                   & (-0.0069) &  (0.4816) &  (0.0057) \\
\hline
\textbf{22} & Nd$^{3+}$ & $f^{3}$       &$|^{4}F_{7/2}\rangle$    & -0.9648    & 8     & 13383   & 6        &  0.4650  &  0.0533  &  -0.4233  \\
  &           &         &                       & & & &                                                   & (0.4601) &  (0.0537) &  (-0.4552) \\
\hline
\textbf{23} & Nd$^{3+}$ & $f^{3}$       &$|^{4}S_{3/2}\rangle$    & 0.9719    & 9     & 13429   & 5        &  0.0362  &  0  &  0  \\
  &           &         &                       & & & &                                                   & (0) &  (0) &  (0) \\
\hline
\end{tabular}
  \label{onecolumntable}
  \end{center}
\end{table*}


\clearpage

\begin{table*}[b]
\noindent Table 1 - cont. \\
\begin{center}
  \begin{tabular}[htbp]{@{}cccclcrcccc@{}}
\hline \textbf{No}.    &  \multicolumn{2}{c}{\textbf{RE ion}}
&\multicolumn{5}{c}{\textbf{Electron eigenstate data}}
    &  \multicolumn{3}{c}{\textbf{Multipolar asphericity}} \\
    & Ion          &     conf. & Up.Comp.
    & Ampl. & no.& E & n  & $A_{2}(\Psi)$   &  $A_{4}(\Psi)$  & $A_{6}(\Psi)$  \\
    \hline
\textbf{41} & Sm$^{3+}$ & $f^{5}$       &$|^{6}H_{15/2}\rangle$    & -0.9782    & 7    & 6431   & 8        &  1.5794  &  0.9086  &  -0.8624  \\
  &           &         &                       & & & &                                                   & (1.6095) &  (0.9134) &  (-0.9000) \\
\hline
\textbf{42} & Sm$^{3+}$ & $f^{5}$       &$|^{6}F_{3/2}\rangle$    & 0.9779    & 8    & 6548   & 10        &  0.0565  &  0  &  0  \\
  &           &         &                       & & & &                                                   & (0.0596) &  (0) &  (0) \\
\hline
\textbf{43} & Sm$^{3+}$ & $f^{5}$       &$|^{6}F_{5/2}\rangle$    & -0.9719    & 9    & 7052   & 20        &  0.0033  &  -0.1068  &  0  \\
  &           &         &                       & & & &                                                   & (0.0130) &  (-0.1260) &  (0) \\
\hline
  \end{tabular}
  \label{onecolumntable}
\end{center}
\end{table*}


\clearpage

\begin{table*}[b]
\noindent Table 1 - cont. \\
\begin{center}
  \begin{tabular}[htbp]{@{}cccclcrcccc@{}}
\hline \textbf{No}.    &  \multicolumn{2}{c}{\textbf{RE ion}}
&\multicolumn{5}{c}{\textbf{Electron eigenstate data}}
    &  \multicolumn{3}{c}{\textbf{Multipolar asphericity}} \\

    & Ion          &     conf. & Up.Comp.
    & Ampl. & no.& E & n  & $A_{2}(\Psi)$   &  $A_{4}(\Psi)$  & $A_{6}(\Psi)$  \\
    \hline
\textbf{61} & Tb$^{3+}$ & $f^{8}$       &$|^{7}F_{0}\rangle$    & 0.9736    & 7   & 5717   & 7            &  0  &  0  &  0  \\
  &           &         &                       & & & &                                                   & (0) &  (0) &  (0) \\
\hline
\textbf{62} & Dy$^{3+}$ & $f^{9}$       &$|^{6}H_{15/2}\rangle$    & -0.9708    & 1   & 0   & 8            & -1.5745  &  -0.9047  &  0.8503  \\
  &           &         &                       & & & &                                                   & (-1.6095) &  (-0.9134) &  (0.9000) \\
\hline
\textbf{63} & Dy$^{3+}$ & $f^{9}$       &$|^{6}H_{13/2}\rangle$    & -0.9830    & 2   & 3432   & 7            & -1.2254  &  -0.3172  &  -0.3992  \\
  &           &         &                       & & & &                                                   & (-1.2375) &  (-0.3128) &  (-0.4146) \\
\hline

  \end{tabular}
  \label{onecolumntable}
\end{center}
\end{table*}


\clearpage

\begin{table*}[b]
\noindent Table 1 - cont. \\
\begin{center}
  \begin{tabular}[htbp]{@{}cccclcrcccc@{}}
\hline \textbf{No}.    &  \multicolumn{2}{c}{\textbf{RE ion}}
&\multicolumn{5}{c}{\textbf{Electron eigenstate data}}
    &  \multicolumn{3}{c}{\textbf{Multipolar asphericity}} \\
       & Ion          &     conf. & Up.Comp.
    & Ampl. & no.& E & n  & $A_{2}(\Psi)$   &  $A_{4}(\Psi)$  & $A_{6}(\Psi)$  \\
 \hline
\textbf{81} & Ho$^{3+}$ & $f^{10}$       &$|^{3}L_{9}\rangle$    & 0.9462    & 21   & 28805   & 2            & 1.2573  &  1.0060  & -0.6106  \\
  &           &         &                       & & & &                                                   & (1.1349) &  (1.1236) &  (-0.8368) \\
\hline
\textbf{82} & Er$^{3+}$ & $f^{11}$       &$|^{4}I_{15/2}\rangle$    & 0.9852    & 1   & 0   & 3            & 0.6798  &  0.6933  & 1.7442  \\
  &           &         &                       & & & &                                                   & (0.6438) &  (0.6850) &  (1.7999) \\
\hline
\textbf{83} & Er$^{3+}$ & $f^{11}$       &$|^{4}I_{13/2}\rangle$    & 0.9955    & 2   & 6514   & 3            & 0.5670  &  0.4715  & 0.6150  \\
  &           &         &                       & & & &                                                   & (0.5569) &  (0.4691) &  (0.6217) \\
\hline
  \end{tabular}
  \label{onecolumntable}
\end{center}
\end{table*}


\clearpage

\begin{table*}[htbp]
\noindent Table 1 - cont. \\
\begin{center}
  \begin{tabular}[htbp]{@{}cccclcrcccc@{}}
\hline \textbf{No.}    &  \multicolumn{2}{c}{\textbf{RE ion}}
&\multicolumn{5}{c}{\textbf{Electron eigenstate data}}
    &  \multicolumn{3}{c}{\textbf{Multipolar asphericity}} \\
    & Ion          &     conf. & Up.Comp.
    & Ampl. & no.& E & n  & $A_{2}(\Psi)$   &  $A_{4}(\Psi)$  & $A_{6}(\Psi)$  \\
    \hline
\textbf{101} & Tm$^{3+}$ & $f^{12}$       &$|^{1}I_{6}\rangle$    & 0.9956    & 10   & 34201   & 2            & 3.0185  & -1.4165  & 0.4414  \\
  &           &         &                       & & & &                                                   & (3.0318) &  (-1.4375) &  (0.4554) \\
\hline
\textbf{102} & Tm$^{3+}$ & $f^{12}$       &$|^{3}P_{1}\rangle$    & 1    & 11   & 34636   & 1            & 0.5477  & 0  & 0  \\
  &           &         &                       & & & &                                                   & (0.5477) &  (0) &  (0) \\
\hline
\textbf{103} & Tm$^{3+}$ & $f^{12}$       &$|^{3}P_{2}\rangle$    & -0.7047    & 12   & 37213   & 3            & -0.9160  & -0.5731  & 0  \\
  &           &         &                       & & & &                                                   & (-0.8366) &  (0) &  (0) \\
\hline
  \end{tabular}
\end{center}
\end{table*}


\clearpage
\begin{table*}[htbp]
  \caption{Multipole characteristics of the RE$^{+3}$ ion eigenstates (selected from
  Table 1) distinguished by the strongest (the upper half) and the
  weakest (the lower half) $A=(A_{2}^{2}+A_{4}^{2}+A_{6}^{2})^{1/2}$}

\begin{center}
  \begin{tabular}[htbp]{@{}cccc@{}}
\hline
No. in & RE& The upper state & A\\
Table 1  & ion & $^{2S+1}L_{J}$ & \\
\hline
13              & Pr$^{3+}$    &           $^{1}I_{6}$          & 3.3784 \\
101             & Tm$^{3+}$     &           $^{1}I_{6}$          & 3.3634 \\
5               & Pr$^{3+}$     &           $^{3}H_{6}$          & 2.1217 \\
92              & Tm$^{3+}$     &           $^{3}H_{6}$          & 2.1169 \\
18              & Nd$^{3+}$     &           $^{4}I_{15/2}$       & 2.0160 \\
41              & Sm$^{3+}$     &           $^{6}H_{15/2}$       & 2.0159 \\
62              & Dy$^{3+}$     &           $^{6}H_{15/2}$       & 2.0051 \\
82              & Er$^{3+}$     &           $^{4}I_{15/2}$       & 1.9962 \\
33              & Pm$^{3+}$     &           $^{5}I_{8}$          & 1.8909 \\
74              & Ho$^{3+}$    &           $^{5}I_{8}$          & 1.8301 \\
\hline
23              & Nd$^{3+}$     &           $^{4}S_{3/2}$        & 0.0361 \\
80              & Ho$^{3+}$     &           $^{5}S_{2}$          & 0.0500 \\
72              & Dy$^{3+}$     &           $^{6}F_{3/2}$        & 0.0502 \\
42              & Sm$^{3+}$     &           $^{6}F_{3/2}$        & 0.0565 \\
43              & Sm$^{3+}$     &           $^{6}F_{5/2}$        & 0.1068 \\
71              & Dy$^{3+}$     &           $^{6}F_{5/2}$        & 0.1503 \\
87              & Er$^{3+}$     &           $^{4}S_{3/2}$        & 0.1688 \\
34              & Pm$^{3+}$     &           $^{5}F_{1}$          & 0.2180 \\
25              & Nd$^{3+}$     &           $^{2}H(2)_{11/2}$    & 0.2402 \\
28              & Nd$^{3+}$     &           $^{2}P_{3/2}$        & 0.2818 \\
\hline
  \end{tabular}
\end{center}
\end{table*}




\clearpage
\begin{table*}[htbp]
  \caption{Multipole characteristics of the RE$^{+3}$ ion eigenstates (selected from
  Table 1) distinguished by the strongest $|A_{k}|$}

\begin{center}
\begin{tabular}[htbp]{@{}cccc@{}}
\hline
No. in & RE& The upper state & $A_{2}$\\
Table 1  & ion & $^{2S+1}L_{J}$ & \\
\hline
13              & Pr$^{3+}$     &           $^{1}I_{6}$          & -3.0273 \\
101             & Tm$^{3+}$     &           $^{1}I_{6}$          & 3.0185 \\
41               & Sm$^{3+}$     &           $^{6}H_{15/2}$          & 1.5794 \\
62              & Dy$^{3+}$     &           $^{6}H_{15/2}$          & -1.5745 \\
92              & Tm$^{3+}$     &           $^{3}H_{6}$       & 1.5291 \\
5              & Pr$^{3+}$     &           $^{3}H_{6}$       & -1.5204 \\
\hline
  \end{tabular}
\end{center}
\end{table*}


\clearpage
\begin{table*}[htbp]
  \caption{Multipole characteristics of the RE$^{+3}$ ion eigenstates (selected from
  Table 1) distinguished by the weakest $|A_{k}|$}

\begin{center}
  \begin{tabular}[htbp]{@{}cccc@{}}
\hline
No. in & RE& The upper state & $A_{2}$\\
Table 1  & ion & $^{2S+1}L_{J}$ & \\
\hline
43             & Sm$^{3+}$     &           $^{6}F_{5/2}$          & 0.0033 \\
85             & Er$^{3+}$     &           $^{4}I_{9/2}$          & 0.0067 \\
80               & Ho$^{3+}$     &           $^{5}S_{2}$          & 0.0188 \\
26              & Nd$^{3+}$     &           $^{4}G_{5/2}$          & 0.0321 \\
71              & Dy$^{3+}$    &           $^{6}F_{5/2}$       & -0.0340 \\
23              & Nd$^{3+}$     &           $^{4}S_{3/2}$       & 0.0362 \\
\hline
  \end{tabular}
\end{center}
\end{table*}


\clearpage
\begin{table*}[htbp]
  \caption{Multipole characteristics of the RE$^{+3}$ ion eigenstates (selected from
  Table 1) distinguished by the largest $|A_{k}|/A$}

\begin{center}
  \begin{tabular}[htbp]{@{}cccc@{}}
\hline
No. in & RE& The upper state & $A_{2}/A$\\
Table 1  & ion & $^{2S+1}L_{J}$ & \\
\hline
14             & Pr$^{3+}$     &           $^{3}P_{2}$          & 0.9917 \\
99             & Tm$^{3+}$     &           $^{3}P_{2}$          & -0.9771 \\
6               & Pr$^{3+}$     &           $^{3}F_{2}$          & 0.9659 \\
35              & Sm$^{3+}$     &           $^{6}H_{5/2}$          & 0.9538 \\
64              & Dy$^{3+}$    &           $^{6}H_{11/2}$       & -0.9528 \\
69             & Dy$^{3+}$     &           $^{6}H_{5/2}$       & -0.9488 \\
\hline
  \end{tabular}
\end{center}
\end{table*}


\clearpage
\begin{table*}[htbp]
  \caption{Multipole characteristics of the RE$^{+3}$ ion eigenstates (selected from
  Table 1) distinguished by the smallest $|A_{k}|/A$}

\begin{center}
  \begin{tabular}[htbp]{@{}cccc@{}}
\hline
No. in & RE& The upper state & $A_{2}/A$\\
Table 1  & ion & $^{2S+1}L_{J}$ & \\
\hline
85             & Er$^{3+}$     &           $^{4}I_{9/2}$          & 0.0000 \\
9             & Pr$^{3+}$     &           $^{1}G_{4}$          & -0.0300 \\
26               & Nd$^{3+}$     &           $^{4}G_{5/2}$          & 0.0671 \\
88              & Er$^{3+}$     &           $^{2}H(2)_{11/2}$          & 0.0970 \\
93              & Tm$^{3+}$    &           $^{3}F_{4}$       & -0.1625 \\
51             & Eu$^{3+}$     &           $^{7}F_{4}$       & 0.1664 \\
\hline
  \end{tabular}
\end{center}
\end{table*}

\end{document}